\documentclass[aps,prl,twocolumn,amsmath,amssymb,superscriptaddress]{revtex4-1}
\usepackage{graphicx}
\usepackage{amssymb}
\usepackage{natbib}
\usepackage{xfrac}
\usepackage{epstopdf}
\usepackage{float}
\usepackage{xspace}
\usepackage{xcolor}
\newcommand{\ceco}{CeCoIn$_5$\xspace}
\newcommand{\laco}{LaCoIn$_5$\xspace}

\newcommand{\smco}{SmCoIn$_5$\xspace}

\newcommand{\ef}{$E_F$\xspace}

\newcommand{\cethree}{Ce$^{3+}$\xspace}
\newcommand{\smthree}{Sm$^{3+}$\xspace}
\newcommand{\smtwo}{Sm$^{2+}$\xspace}
\newcommand{\g}{$\Gamma$\xspace}
\newcommand{\eb}{$E_B$\xspace}

\begin{document}


\title{Flat-band hybridization between $f$ and $d$ states near the Fermi energy of \smco}

\author{David~W.~Tam}
\email{david-william.tam@psi.ch}
\affiliation{Laboratory for Neutron Scattering and Imaging, Paul Scherrer Institut, 5232 Villigen, Switzerland}
\author{Nicola~Colonna}
\affiliation{Laboratory for Materials Simulations, Paul Scherrer Institut, 5232 Villigen, Switzerland}
\affiliation{Laboratory for Neutron Scattering and Imaging, Paul Scherrer Institut, 5232 Villigen, Switzerland}
\affiliation{National Centre for Computational Design of Novel Materials (MARVEL), Paul Scherrer Institut, 5232 Villigen, Switzerland}
\author{Fatima~Alarab}
\affiliation{Photon Science Division, Paul Scherrer Institut, 5232 Villigen, Switzerland}
\author{Vladimir~N.~Strocov}
\affiliation{Photon Science Division, Paul Scherrer Institut, 5232 Villigen, Switzerland}
\author{Dariusz~Jakub~Gawryluk}
\affiliation{Laboratory for Multiscale Materials Experiments, Paul Scherrer Institute, 5232 Villigen, Switzerland}
\author{Ekaterina~Pomjakushina}
\affiliation{Laboratory for Multiscale Materials Experiments, Paul Scherrer Institute, 5232 Villigen, Switzerland}
\author{Michel~Kenzelmann}
\email{michel.kenzelmann@psi.ch}
\affiliation{Laboratory for Neutron Scattering and Imaging, Paul Scherrer Institut, 5232 Villigen, Switzerland}

\date{\today}

\begin{abstract}

We present high-quality angle-resolved photoemission (ARPES) and density functional theory calculations (DFT+U) of \smco.
We find broad agreement with previously published studies of \laco and \ceco \cite{CXNJ17,CLVK19}, confirming that the Sm $4f$ electrons are mostly localized.
Nevertheless, our model is consistent with an additional delocalized Sm component, stemming from hybridization between the $4f$ electrons and the metallic bands at ``hot spot'' positions in the Brillouin zone.
The dominant hot spot, called $\gamma_Z$, is similar to a source of delocalized $f$ states found in previous experimental and theoretical studies of \ceco \cite{CXNJ17,JDAZ20}.
In this work, we identify and focus on the role of the Co $d$ states in exploring the relationship between heavy quasiparticles and the magnetic interactions in \smco, which lead to a magnetically ordered ground state from within an intermediate valence scenario \cite{IHFS06,PJWR18,sm115delocalization}.
Specifically, we find a globally flat band consisting of Co $d$ states near $E=-0.7$ eV, indicating the possibility of enhanced electronic and magnetic interactions in the ``115'' family of materials through localization in the Co layer, and we discuss a possible origin in geometric frustration.
We also show that the delocalized Sm $4f$ states can hybridize directly with the Co $3d_{xz}$/$3d_{yz}$ orbitals, which occurs in our model at the Brillouin zone boundary point $R$ in a band that is locally flat and touches the Fermi level from above.
Our work identifies microscopic ingredients for additional magnetic interactions in the ``115'' materials beyond the RKKY mechanism, and strongly suggests that the Co $d$ bands are an important ingredient in the formation of both magnetic and superconducting ground states.
\end{abstract}
\maketitle


The origin of heavy quasiparticles with $f$ electron character is of fundamental interest due to their participation in unconventional superconductivity.
In heavy fermion metals with a valence band of metallic $s$, $p$, and $d$ electrons, the conduction electrons can turn into heavy quasiparticles by hybridizing with the localized $f$ electrons.
Hybridization of this type, involving the formation of entangled Kondo singlets that screen the $f$ magnetic moments, has been observed in a variety of Ce-based intermetallics \cite{ShHK07,FFSN06,KBIG07,WhTM15,PGGK16,CXNJ17,JDAZ20,KFNM21,MECW22,XuCZ22}.
Microscopically, the Kondo hybridization mechanism depends on the details of the electronic wavefunctions, and it has been argued that hybridization is driven by the shape of the $f$ orbital wavefunctions as they are determined by the crystal electric field \cite{WHHK10,WSHS15}.

The single-ion properties of \smthree ($4f^5$, $S=\sfrac{5}{2}$, $L=5$, $J=\sfrac{5}{2}$) make it an analogue of \cethree in \ceco ($4f^1$, $S=\sfrac{1}{2}$, $L=3$, $J=\sfrac{5}{2}$) with the same nominal $J$, while exhibiting similar single-ion properties, including the crystal field environment and atomic spin-orbit coupling.
The case of \smco presents with predominantly localized $f$ electrons, as well as an additional delocalized $f$ component \cite{sm115delocalization}, similar to the case of \ceco.
However, unlike \ceco, the ground state of \smco exhibits several antiferromagnetic (AF) ordered phases, with three phase transitions observed in specific heat at $T_{N,1} \approx 11$ K, $T_{N,2} \approx 8$ K, and $T_{N,3} \approx 6$ K \cite{IHFS06,PJWR18}.
Magnetic ordering in rare earth intermetallics is expected to be dominated by the RKKY mechanism \cite{CJKP14,WhTM15,MGMY19}.
In case of an additional Kondo hybridization, the RKKY interactions are reduced by the Kondo screening cloud, which typically prohibits the formation of long-range magnetic order.
Therefore, the microscopic differences between \ceco and \smco, especially changes in their most dominant magnetic interactions, may hold the key to understanding how magnetic ground states can emerge from within a system exhibiting Kondo screening.

\begin{figure*}[tbph]
\includegraphics[scale=.48]{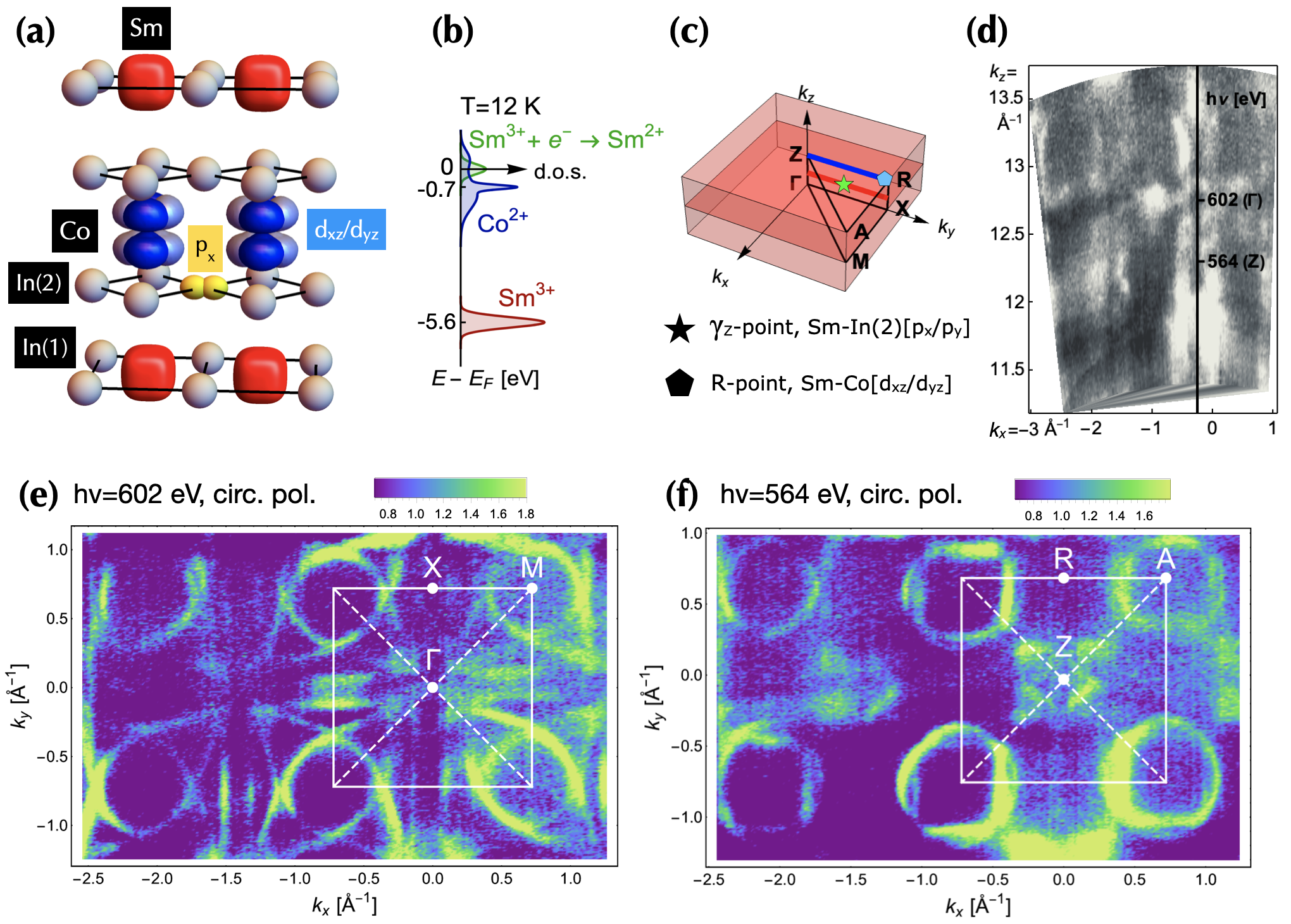}
\caption{
(a) Tetragonal unit cell of \smco, showing the position and orientation of the In(2) $p_x$ orbital (yellow) and Co $d_{xz}/d_{yz}$ orbitals (light and dark blue, respectively).
(b) Sketch of the density of states of \smco relative to \ef.
(c) Conventional labels within the $P4/mmm$ Brillouin zone, highlighting two positions of interest $\gamma_Z$ ($k=[0.25,0,0.2]$) and $R$ ($k=[0.5,0,0.5]$).
(d) ARPES intensity at $T=12$ K, parallel to the \g-$X$ direction, while varying photon energy $h\nu$, demonstrating the coupling of photon energy to the out-of-plane $k_z$ direction and identifying the energies corresponding to two high-symmetry planes along the $k_z$ axis.
(e) Fermi surface of \smco in the \g-$X$-$M$ plane ($k_z$=0) using circularly polarized light with photon energy $h\nu=602$ eV, and (f) $Z$-$R$-$A$ plane ($k_z=\pi$) using circularly polarized light with photon energy $h\nu=564$ eV.
}
\label{fig-fs}
\end{figure*}

To develop a microscopic picture of the Kondo screening environment in \smco, we conducted ARPES measurements at T=12 K and built an electronic structure model using DFT+U calculations.
Our results confirm that \smco is dominated by the \smthree configuration \cite{PJWR18,sm115delocalization} and is highly similar to other ``115'' materials, where the $f$ electrons remain mostly localized and play only a small role in the overall characteristics of the band structure.
Specifically, we find that the $f$ electrons of \smthree sit near a binding energy of $E_B \approx 6$ eV, with good agreement between ARPES and our DFT+U results, showing little hybridization of the \smthree states.
In addition, a second band of Sm states sits above this position by a value $E_B+U$, which happens to nearly overlap the Fermi energy \ef since $U \approx 6$ eV.
We interpret this band loosely as an ``upper Hubbard band'' of Sm, because it comprises a situation where an $s,p,d$ conduction electron tunnels onto the localized \smthree site, leading to an instantaneous \smtwo configuration.
To show that this picture is correct, we also conducted separate DFT calculations within a strictly \smtwo configuration, which confirm that the \smtwo states appear close to the Fermi energy (see supplementary information), although the details of that calculation also show that the \smtwo configuration cannot be the ground state of \smco.
A partial \smtwo character is also in agreement with previous experiments that show a slight admixture of the \smtwo valence state at low temperatures below T=60 K \cite{sm115delocalization}, similar to the intermediate valence state found in a variety of other Sm- and Yb-based materials \cite{HYMA18,KGKZ18,LHHD22}.
Therefore, we believe the origin of the intermediate valence scenario in \smco is captured by our DFT calculations.

\begin{figure*}[tbph]
\includegraphics[scale=.41]{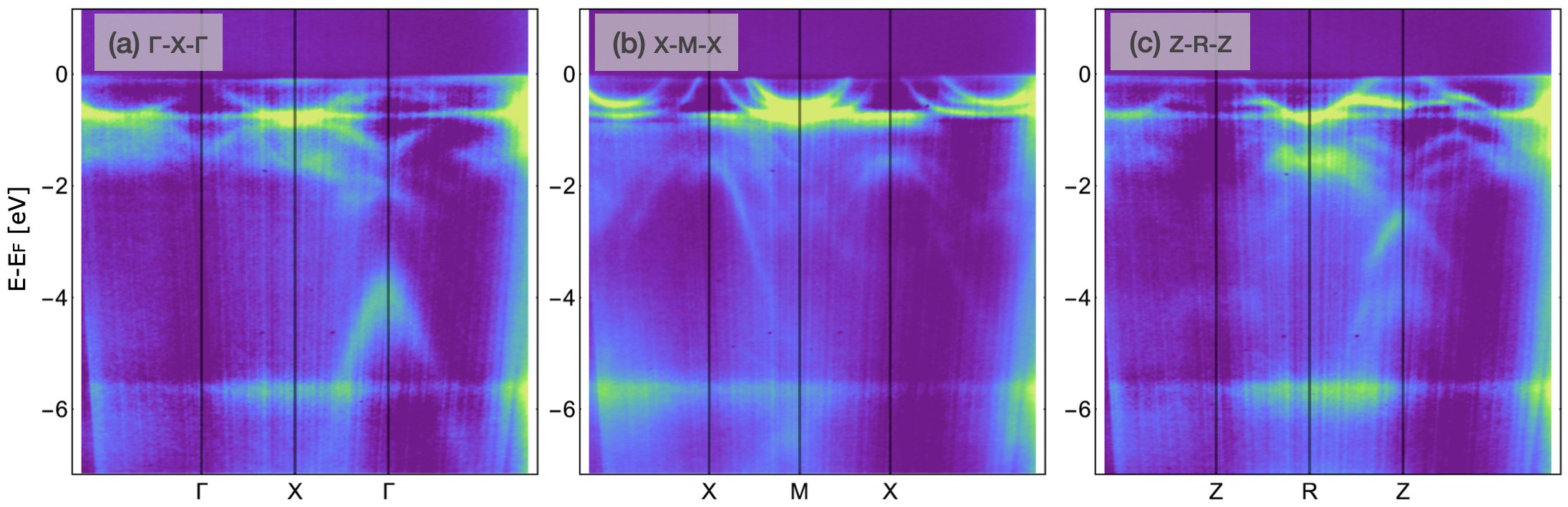}
\caption{
ARPES \eb-$k_x$ maps showing the band structure of \smco.
(a) \g-$X$-\g cut for $E-E_F>$ -7 eV. The data has been partially background-subtracted using a line-derivative method in order to enhance the visibility of the bands.
(b) $X$-$M$-$X$ cut.
(c) $Z$-$R$-$Z$ cut.
}
\label{fig-vb}
\end{figure*}

Looking in more detail, we first show (Fig. 1-3) that the band structure observed by ARPES agrees very well with the DFT+U calculations, giving us a high degree of confidence in our model.
The model reveals the ``hot spots'' in the Brillouin zone where the Sm $f$ states hybridize with the conduction electrons near the Fermi energy \ef, and we find two spots of interest which we label $\gamma_Z=(0.25,0,0.2)$ and $R=(0.5,0,0.5)$ (Fig. 1c; Fig. 4).
The $\gamma_Z$ spot sits midway along the $\Gamma-X$ cut direction with nonzero $k_z=0.2$, which is similar to a large hot spot position previously found in ARPES, DFT, and DMFT studies of \ceco \cite{CXNJ17,JDAZ20,KFNM21}.
In \smco, our model shows a smaller hot spot, which we argue may be related to the different single-ion properties of \cethree and \smthree.

The second ``hot spot'' for $f$ hybridization in \smco is the high-symmetry $R$ position (Fig. 3-4), located at the Brillouin zone boundary along both $k_x$ and $k_z$, which was also previously identified as weakly important for \ceco \cite{JDAZ20} but more important for Sn-doped \ceco \cite{MECW22}, which is in the limit of higher $f$ electron itinerancy (higher \ef).
Our model calculations (Fig. 3-4) unambiguously show that the $R$ point contains only Sm-$f$ and Co-$d$ states, indicating that a direct hybridization mechanism is possible.
Moreover, orbitally resolving the model (Fig. 5) shows that the $R$ point contains only the Co $3d_{xz}$ orbital (or the equivalent $3d_{yz}$ orbital at the $R$ point rotated 90 degrees about the origin).
Since the $d$-$f$ band is highly flat at the $R$ point, it is likely that a large van Hove peak of Sm/Co states suddenly appears in the density of states, pinning the Fermi energy at the bottom of this band.
The appearance of this singularity exactly at \ef suggests the $R$ point plays a role in the ground state of \smco, even though we do not observe it in our experiments.
More importantly, due to the direct $f$-$d$ hybridization within this band, our model shows generally that magnetic interactions of the Co $d$ states can directly influence the magnetic degree of freedom of the heavy quasiparticles, whether through interactions at the $R=(0.5,0,0.5)$ wavevector or elsewhere.

Finally, our model shows unambiguously that a globally flat band appearing at $E=-0.7$ eV in \smco, as well as in LaCoIn$_5$ and \ceco \cite{CXNJ17,CLVK19}, consists of Co $d$ states.
Since our experiments and calculations both indicate that the band is globally flat, we suggest that the flatness of this band arises from a geometric frustration mechanism in the quasi-2D Co-In(2) layer.
Specifically, we point out several square lattice models, including the checkerboard and Lieb lattices, that could apply to a subgroup of the Co $d$ orbitals (see Discussion).
Our results show microscopically that this flat band, where electronic and magnetic correlations are expected to be enhanced, can influence the heavy quasiparticles, either through direct exchange (such as at the R point) or via superexchange across the In(2) sites.
We argue then that geometric frustration and the resulting flat $d$ band may help explain the appearance of magnetic order within a system where the $f$ states favor a delocalized ground state and, more broadly, that it might play an important (and until now overlooked) role in understanding magnetism and superconductivity in the ``115'' materials.

\begin{figure*}[tbph]
\includegraphics[scale=.52]{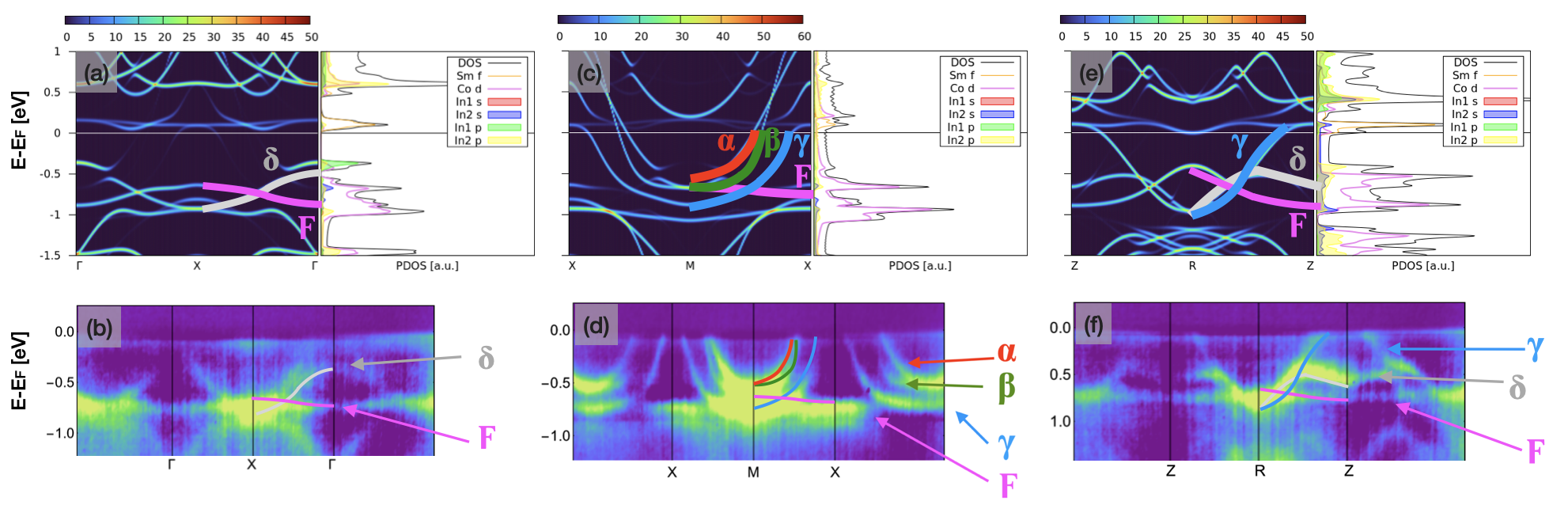}
\caption{
Comparison between DFT+U model calculations and ARPES within 1 eV of \ef.
(a) DFT calculated band structure and density of states along the \g-X-\g cut, and (b) comparison with ARPES data that has been enhanced using the background subtraction method of Fig. 2.
The lines that are drawn in by hand over the DFT panel are copied directly onto the ARPES panel below, signifying good agreement, and the colored labels specify the band names discussed in the text.
Note that the colors of these bands are not related to the colors indicated in the density of states plots on the right-hand side of each DFT plot.
(c-d) X-M-X cut.
(e-f) Z-R-Z cut.
}
\label{fig-vb-annot}
\end{figure*}

\section{Results and discussion}

\subsection{Fermi surface and band structure of \smco}

Fig. 1 shows the Fermi surface of \smco in the $\Gamma$-X-M and Z-R-A planes.
ARPES intensity maps were taken as a function of the in-plane momentum ($k_x$,$k_y$) at the Fermi energy \ef using two photon energies corresponding to the $\Gamma$ and $Z$ points (Fig. 1d).
Using the standard notation used for ``115''s, e.g. of Ref. \cite{CLVK19}, we observe three electron-like bands at the M point ($\alpha$, $\beta$, and $\gamma$).
The $\alpha$ and $\beta$ sheets are visible around the M point with circular and cloverleaf shapes, respectively.
The $\gamma$ sheet is visible as parallel lines around the X point, as well as in some of the features close to $\Gamma$.
In a DMFT study of \ceco, the $\gamma$ sheet structure near X was shown to exhibit an in-plane curvature that pinches the pocket to a narrow point directly on the line between $\Gamma$ and X.
In \smco, we observe that the $\gamma$ sheet is highly parallel, with no in-plane curvature visible until the intersection of the $\beta$ and $\gamma$ pockets very close to the $\Gamma$ point.
The absence of such a pinch point suggests that the Sm $f$ electrons do not enter the Fermi surface here, whereas they do in \ceco \cite{JDAZ20,KFNM21}.
Instead, our model shows the $f$ electrons in \smco only enter the Fermi surface at the $\gamma_Z$ and $R$ positions, as we discuss below.
The absence of an in-plane hot spot might imply a more three-dimensional character to the Kondo screening in \smco.

In Fig. 2, we show ARPES intensity maps ($E_B$-$k_x$) showing the band structure of \smco along high-symmetry cuts between \ef and binding energy $E_B=7$~eV.
The flat band near $E_B=$-5.6 eV is from \smthree, and its flatness reflects the predominantly localized nature of \smthree, wherein the $f$ electrons cannot hybridize strongly.
The band structure near \ef is accurately captured by our DFT calculations in minute detail, as shown in Fig. 3.
In Fig. 3(a,c,e) we zoom in to the band structure near \ef along the same high-symmetry cuts adopted in Fig. 2.
Around the $M$ point, shown in Fig. 3(c-d), electronic-like $\alpha$, $\beta$, and $\gamma$ sheets are highly consistent with the band structure of \laco and \ceco \cite{CXNJ17,CLVK19}.
At the Brillouin zone boundary at $Z$, shown in Fig. 3(e-f), the $\gamma$ sheet creates an open pocket at \ef near $Z$, while the lower $\delta$ sheet is closed at both $\Gamma$ and Z and exhibits an inversion from concave down to concave up between $\Gamma$ and $Z$.
In a previous ARPES study of \ceco \cite{KFNM21}, it was reported that depending on the surface termination, some bulk features could appear in different layers of the experimental data.
In our data, we observe such an extra band in the \g-$X$-\g cut (Fig. 3[b], about 0.2 eV above the $\delta$ band and forming an open pocket at $\Gamma$), which closely resembles the the $\gamma$ sheet that is found by our bulk DFT calculations only along the $Z$-$R$-$Z$ cut (Fig. 3[e-f]).
Since this is precisely the same termination state proposed for a Ce-In(1) surface in \ceco (cf. Fig. 2a of Ref. \cite{KFNM21}), we attribute this feature in our data to a 2D termination state containing the Sm-In(1) crystallographic layer.

In all cuts shown in Fig. 2 and 3, we observe the presence of a very strong and flat band near $E-E_F=-0.7$~eV, which we label with the archaic Greek letter $\digamma$ (digamma).
A similar flat band was also observed in \laco and \ceco \cite{CXNJ17,CLVK19}, but it was not clearly determined if this band arises from an ionization state of the localized rare earth ion, or if it is part of the $p$ and $d$ metallic bands.
From our DFT calculations on \smco, we can confirm that the origin of this flat band is from Co $3d$ orbitals, as can be seen in the density of states in Fig. 3(a,c,e), shown on the right side of each subpanel.
By comparing the experimental ARPES data to the calculations in Fig. 3(b,d,f), we note several unusual properties of the $\digamma$ band.
Firstly, its strong intensity (on par with the Sm flat band at $E$-\ef=-5.6 eV), overall flatness throughout the Brillouin zone, and lack of any observable hybridization points (anti-crossings) with any other metallic band, suggests that it represents a significant and relatively isolated subsystem of Co $d$ states.
Second, we note the possibility that the band is even flatter than the model calculations predict, which would indicate the presence of electronic correlations in the $d$ band beyond the DFT+U level of approximation.
The flat band may arise from geometric frustration in a sublattice of Co orbitals, as we show in the following sections, and therefore give rise to additional magnetic interactions.
Moreover, since the Co $d$-shell is unfilled and therefore has a magnetic degree of freedom, the $\digamma$ band can additionally couple magnetically to the entangled Kondo states, which are magnetic singlets. 
In general, the presence of the $\digamma$ band shows that strong correlations in the $d$ states can directly influence the magnetic properties of the heavy quasiparticles near \ef.

\begin{figure*}[tbph]
\includegraphics[scale=.44]{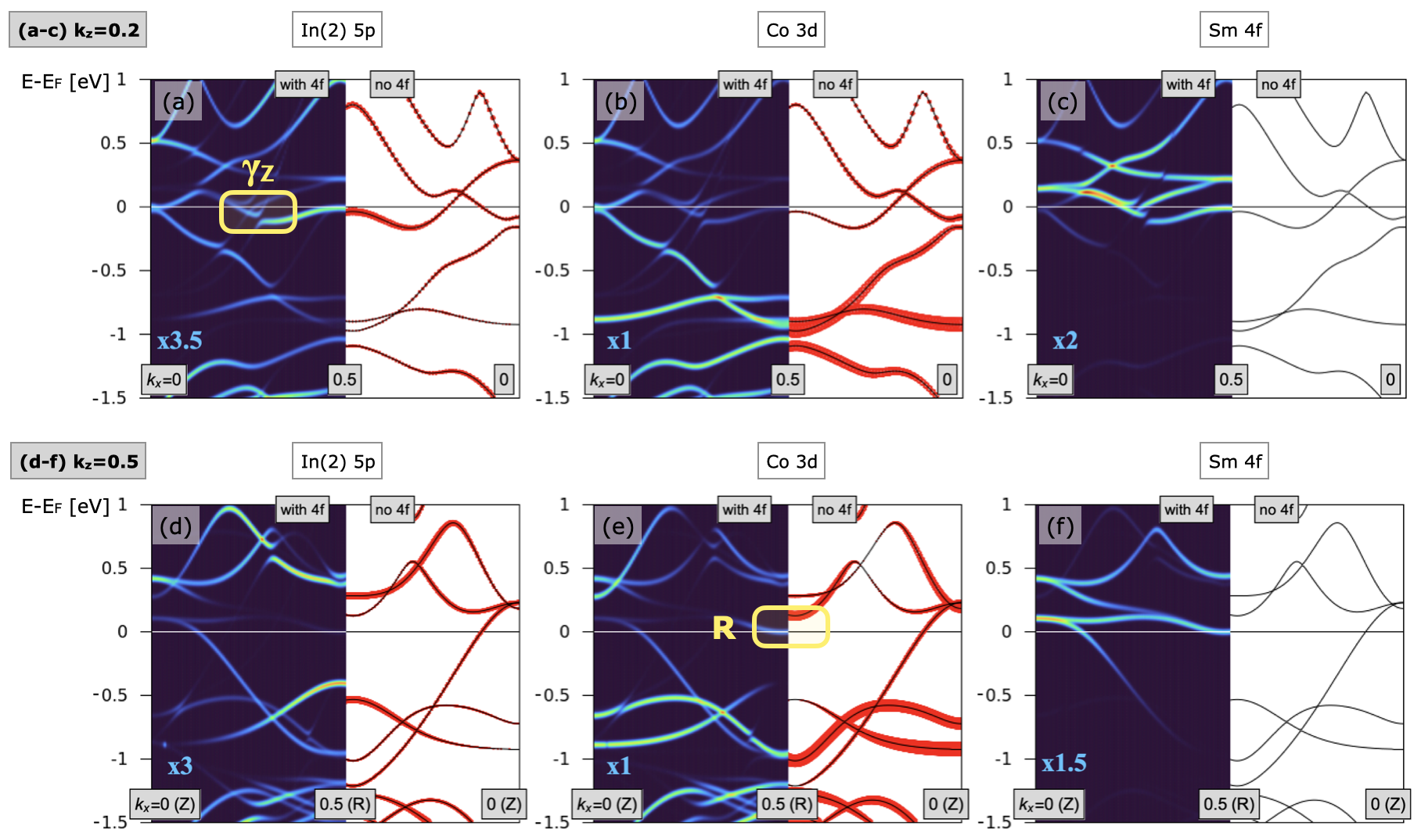}
\caption{
Atomically resolved DFT calculated band structure of \smco parallel to the $\Gamma$-$X$-$\Gamma$ high-symmetry cut, with (a-c) $k_z=0.2$ and (d-f) $k_z=0.5$ ($Z$-$R$-$Z$).
Panels (a,d) show spectral weight of the In(2) site $5p$ orbitals, (b,e) show Co $3d$, and (c,f) show Sm $4f$.
Each sub-panel shows our calculations for \smco (left side) along with a nonmagnetic calculation with the $4f$ electrons frozen in the core (right side).
In the left sub-panels, the intensity of the bands is proportional to the specific atomic orbital content of each band, and is multiplied by the scale factor written in light blue at the bottom of the plot.
For example, in panel (a), the band colors for In(2) 5p have been multiplied by a factor 3.5 compared to the density of Co $3d$ states in panel (b).
For the nonmagnetic calculations in the right sub-panels, the specific orbital content is proportional to the thickness of the red lines superimposed on the bands.
The differences between the magnetic and nonmagnetic calculations is evidenced by slight changes to the position of the bands due to the presence of Sm $f$ electrons.
Two ``hot spots'' of interest at $\gamma_Z$ and $R$ (highlighted in panels [a] and [e]) show a large density of Sm $4f$ states, as shown in panel (c) and panel (f).
In the supplementary information, we additionally show similar plots with projections into the In(1) 5s, In(1) 5p, and In(2) 5s orbitals.
}
\label{fig-dft}
\end{figure*}

\subsection{Hybridization points between $f$ states and metallic bands}

To understand the details of the hybridization between Sm $f$ electrons and the other bands, we projected our DFT calculations into the basis of different atomic shells and studied the ``hot spots'' near the Fermi energy that contain Sm states.
In Fig. 4, we show projections along two high symmetry cuts parallel to the $\Gamma$-$X$-$\Gamma$ direction, for $k_z=0.2$ (a-c) and $k_z=0.5$ (d-f), both with $k_y=0$.
In panel (a), we use a yellow box to highlight the $\gamma_Z$ hot spot position $k=(0.25,0,0.2)$, which contains significant Sm weight (panel [c]).
We find that the In(2) $5p$ orbitals dominate the shallow band extending from $\gamma_Z$ up to $k_x=0.5$ (the $X$ point), while the $\gamma_Z$ position itself contains a mixture of In(1) and In(2) states, with the In(2) $5p$ states providing the largest contribution (contributions from the other orbitals In[1] 5s, In[1] 5p, and In[2] 5s are shown in the supplementary information).
The existence of Sm states near the $X$ point below \ef represents another difference between \ceco and \smco.

From both the model calculations and experimental data, the $\gamma_Z=(0.25,0,0.2)$ hot spot in \smco is smaller and more isolated compared to \ceco.
Specifically, we find no evidence that the $f$ states at $\gamma_Z$ are distributed over the larger $Z$-centered Fermi surface pocket at \ef (see supplementary figure 4(g); cf. Ref. \cite{JDAZ20}), and we find no significant $f$ contributions in the middle of the $Z$-$R$ cut (Fig. 4(f); cf. Ref. \cite{ShHK07,JDAZ20}).
Instead, our DFT+U model shows that the states near the $Z$ point are dominated by the conduction electrons, and we find no experimental evidence of any deviations from the model along the $Z-R-Z$ cut that would show the presence of $f$ states.
However, we have retained the notation $\gamma_Z$ in this work to refer to the hybridization hot spot $(0.25,0,0.2)$, and to avoid confusion with the Fermi surface $\gamma$ sheet.
The observation of a small $\gamma_Z$ hot spot in \smco could be related to different hybridization conditions between the relatively anisotropic \cethree ground state wavefunctions \cite{WHHK10,WSHS15}, compared to the highly isotropic wavefunctions of \smthree \cite{sm115delocalization}.
The lower anisotropy would also explain why a DFT+U approach results in a very realistic model for \smco without the need to take into fully account for the effects of the small crystal field splitting of the lowest $f$ manifold, which is beyond a mean field approach.
Our model shows that the $\gamma_Z$ position hybridizes slightly with the $5s$ and $5p$ orbitals of both In sites, but is dominated by \smthree spectral weight.
Since the In(2) site was also shown to have a strong effect on the ground state superconductivity in Ce-based ``115'' compounds \cite{ShHK07,WSHS15}, we show the In(2) character in Fig. 4 while noting that there is an admixture of other orbitals at $\gamma_Z$.
Within the In(2) atom, we find the most important contribution is from the $5p_x$ orbital (or with the equivalent $5p_y$ orbital at the position rotated 90 degrees about the origin), as is expected from the shape of the $5p_x$ orbitals.

Turning to the hot spot at the $R$ point $k=(0.5,0,0.5)$, in Fig. 4(d-f) we show the presence of a shallow band which appears pinned to the Fermi level and consists of Co $d$ and Sm $f$ states.
This $R$-centered hot spot does not appear to be smoothly connected to the $\gamma_Z$ hot spot, but is instead a separate feature with a distinct set of associated quasiparticles.
In supplementary figure 4, we show constant-energy slices plotting $k_z$ against $k_x$, which show that two minima can be clearly distinguished from one another up to energies at least 20 meV above \ef.
Furthermore, the absence of In(2) states in this band (Fig. 4[d]) shows that the Sm and Co states exhibit magnetic coupling through a direct exchange mechanism, which is relevant not only for this locally flat band at $R$, but also more generally shows that direct exchange can be achieved between the rare earth and transition metal electrons in the ``115'' materials.
Since the $d$+$f$ band at $R$ is very shallow, spanning nearly $\sim$0.2 reciprocal lattice units before deviating significantly from $E$=\ef, the $R$ point allows a flat band of $d$ states to directly hybridize with the heavy $f$ quasiparticles.
Moreover, the local flatness of this band suggests that the van Hove singularity in the density of states is responsible for pinning the Fermi energy to the bottom of this band at $R$.
Therefore, although Sm $f$ states are already present at $\gamma_Z$ at energies below \ef, we find that the hybridized flat band at $R$ is probably touching the Fermi energy.

\begin{figure}[tbph]
\includegraphics[scale=.31]{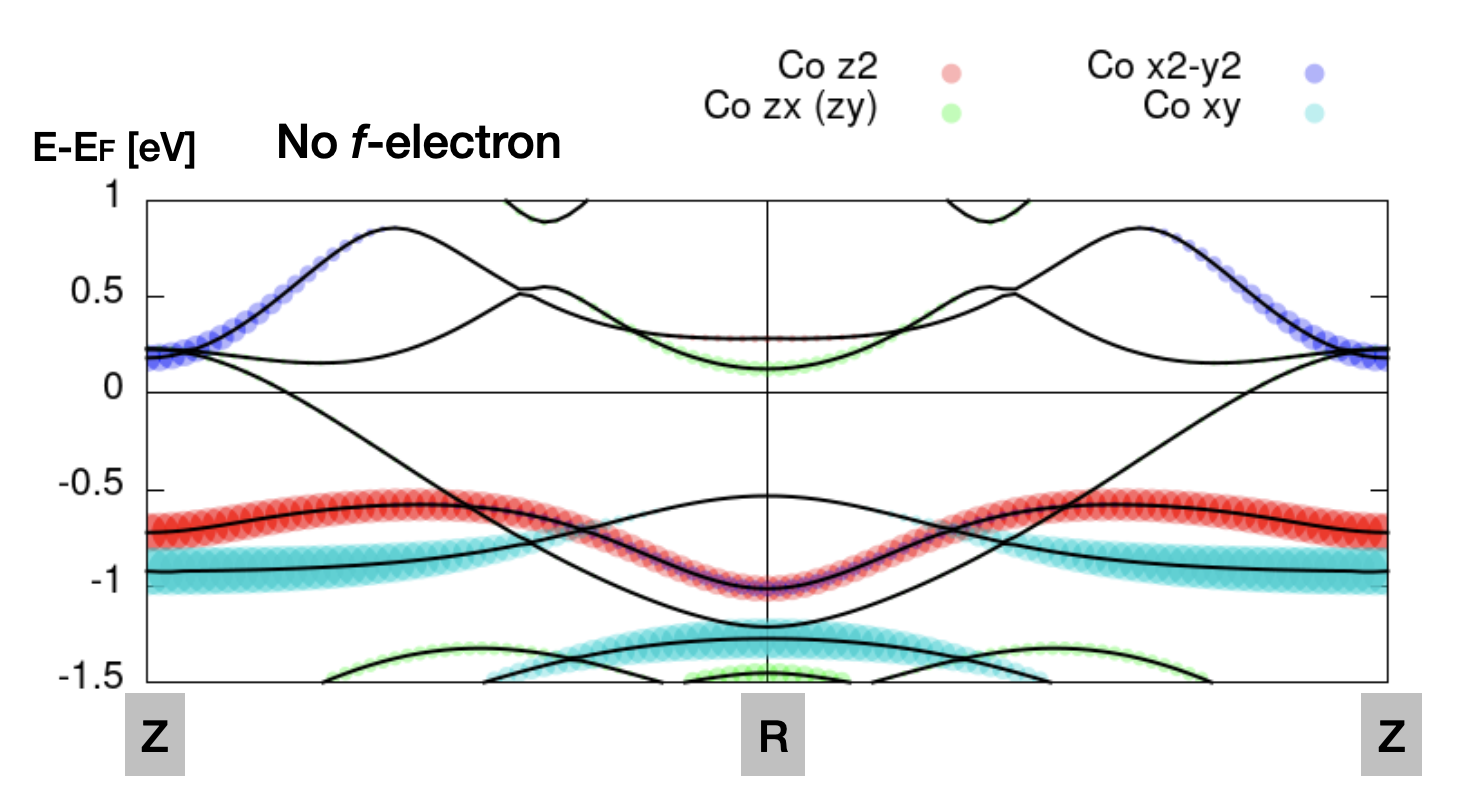}
\caption{
Orbital-resolved Co $3d$ orbitals from a DFT calculation of \smco with $f$ electrons frozen into the core.
The orbitals hybridizing with Sm $f$ states are found to be $d_{x^2-y^2}$ at $Z$, and $d_{xz}$/$d_{yz}$ at $R$.
The flat $\digamma$ band contains $d_{xy}$ and $d_{z^2}$.
}
\label{fig-co}
\end{figure}

To further study how hot spots develop at the $\gamma_Z$ and $R$ hot spots, we conducted separate DFT calculations with the $f$ electrons frozen into the core, shown on the right side of each sub-panel in Fig. 4.
Qualitatively, the two calculations are highly consistent, although one important difference is at the $R$ point, where the Co $d$ band does not dip down low enough to reach \ef in the nonmagnetic case, showing that the interaction between Sm and Co states is the driving mechanism behind creating an interaction between Sm and Co at $R$.
Since the nonmagnetic calculations do not require the use of a magnetic supercell, we also use them to show the proper projection of the In and Co bands onto the orbitals within the crystallographic unit cell illustrated in Fig. 1(a).
In the case of the $\gamma_Z$ position (see supplementary information), we find that the In(2) $5p_x$ orbital dominates the shallow band between $k_x=0.3$ and $0.5$ where the Sm$f$ states occur, while exactly at $\gamma_Z$ it is still the most important of the $s,p,d$ orbitals but by a smaller margin.
Hence, quasiparticles at $\gamma_Z$ exhibit largely Sm-Sm interactions, and could support a Fermi surface nesting instability between neighboring $\gamma_Z$ points---e.g., with wavevector $q=(0.5,0,0.4)$---driven by the Sm electrons.
On the other hand, at the $R$ point, shown in Fig. 5, we find that the $d_{xz}$ orbital completely dominates the $d$+$f$ band at \ef.
This result shows that the hybridization between Sm and Co at $R$ is orbitally selective, such that the interactions affecting the heavy quasiparticles there are related to the symmetry of the Co $d_{xz}$/$d_{yz}$ orbitals.
In the magnetically ordered ground state of \smco \cite{IHFS06,PJWR18}, zone folding and reconstruction of the electronic bands may further increase the weight of the $R$ point below \ef.
Therefore, from a magnetic point of view, the orbital-selective magnetic exchange pathways in \smco involving the $d_{xz}$/$d_{yz}$ orbitals may be an important mechanism competing with Sm-Sm interactions, with consequences for the magnetically ordered ground state of the Sm sublattice.


\subsection{Localization in the Co layer of ``115'' materials}

In our DFT+U calculations, the density of states of the Co $3d$ electrons appears predominantly between \ef and about \eb$=3$ eV, and exhibits a strong flat band ($\digamma$) near $E=-0.7$ eV.
Flat bands in general are hosts of electronic localization (e.g. \cite{TSXX20,PBTK23}), and they have been previously shown to have a strong effect on magnetic ground states in transition metal systems \cite{MCLM00,MATC01,LCZQ18,LLXS19,YZCW19,KFYP20}.
Moveover, since our $\digamma$ band is globally flat throughout the entire Brillouin zone, we are inclined to search for explanations in terms of a robust symmetry-protected mechanism.
To this end, flat bands arising from geometric frustration have been identified in a variety of square or square-like lattices \cite{SGKD11,IHGB14,NoIY15,JPVK16,NuSm20,YOOK20,XuZJ21,FSSL23}, many of which can also host unconventional superconductivity, posing a deeper connection between the $\digamma$ band and the phenomenology of the 115s.

If the $\digamma$ band is globally flat because of geometric frustration, analysis of the 115 crystal structure can elucidate how this occurs.
In the 115 structure, Co atoms occupy a square lattice.
Therefore, one possibility is that the band structure of the 115s hosts a very specific hierarchy of couplings involving the square lattice, such as the checkerboard lattice proposed in \cite{SGKD11}.
On the other hand, as we show in Fig. 1a, the Co site is also coordinated by four In(2) sites just above the plane of Co, and four below.
This Co-In(2) sublattice bears a resemblance to the Lieb lattice, a bipartite square lattice where the central site of the square is coordinated by four neighbors of another species.
Phenomenologically, studies of the Hubbard model on the Lieb lattice have focused on its relationship with both magnetism \cite{Tasa98,NKKP09,NoIY15} and superconductivity \cite{IHGB14,JPVK16}, including as a model for orbital selective superconductivity in the cuprates \cite{YOOK20}.
Since both checkerboard and Lieb lattice models also contain the ingredients for topologically nontrivial band crossings \cite{SGKD11,YaZZ17,FSSL23}, the $\digamma$ band may be important to the 115s in yet another way, since such a crossing was previously proposed in a study of \ceco \cite{SSBP18}.
In general, the possibility of a specific square lattice model or a Lieb lattice model shows that \smco can easily contain the ingredients for geometric frustration in the square-like sublattice of Co $d$ states.

The appearance of geometric frustration does not necessarily imply that the ground state of the frustrated sublattice is the ground state of \smco.
Since the $\digamma$ band in the 115s is not situated at the Fermi energy, its quasiparticles are not thermally activated.
However, since the Co $d$-shell is partially unfilled, the magnetic interactions within the $\digamma$ band can nevertheless directly influence the itinerant quasiparticles.
Also, as we have shown in \smco, direct hybridization can occur between heavy quasiparticles with specific Co $d$ orbitals (in our model, with $d_{xz}$/$d_{yz}$ at the $R$-point), proving that the effects of the Co flat band can be felt directly by the partly delocalized $f$ states.
Therefore, in \ceco, where the delocalized $f$ states participate in a superfluid condensate, interactions arising from the $\digamma$ band of $d$ states might also influence the superconducting pairing mechanism.

\subsection{$f$-Multiplet-Driven Hybridization}

In a DMFT study of \ceco, different multiplets of the Ce $f$ electrons were shown to selectively hybridize with the $\beta$ band between X and M, and with the $\gamma$ band between $\Gamma$ and X \cite{JDAZ20}.
With the much richer multiplet structure of Sm \cite{sm115delocalization,DAKS14}, the hybridization in Sm is more complex, and may contain charge-like degrees of freedom in addition to the typical magnetic Kondo hybridization \cite{sm115delocalization}.
In addition to the spin-orbit coupling, which splits the Sm $f$ levels into multiplets with different $J$, the multi-$f$-electron configuration in Sm also contains Coulomb repulsion between $f$ electrons, a situation not present in the single-$f$-electron case of Ce.
This additional term couples multiplets of different $J$, leading to a vastly larger subspace of $f$ multiplet levels compared to the case of Ce, which is limited to just 14 $f$ states with definite splittings set by the single-ion properties.
Therefore, with Sm it is possible for even a tiny amount of curvature in a nominally flat band to encompass many multiplet states of the Sm ion.
For example, within 0.3 eV of \ef, the ground state of \cethree contains the 14 multiplet levels of different $m_J$, whereas \smthree exhibits nearly 200 different levels that are mixtures of different $J$ and $m_J$ \cite{sm115delocalization}.
Significantly, the various multiplets may have different Kondo coupling strengths that exhibit different hybridization strengths with the itinerant valence band electrons.
This dramatically increases the likelihood of a Kondo-like interaction between at least one of the multiplet states with the valence band, at an energy close enough to \ef to exhibit a Kondo resonance.
The large subspace of $f$ levels in a multi-$f$-electron system may therefore present a natural explanation for why many Sm-based materials exhibit high effective masses \cite{HYMA18}, and is of particular interest in cases such as \smco where the Sm/Co $d$+$f$ band is already locally flat at the $R$ point.
It would therefore be interesting to conduct further photoemission studies on \smco or doped \smco with high resolution, which could possibly show if $f$ multiplet wavefunctions other than the ground state are directly involved in the Kondo coupling, a complexity less likely to be encountered in Ce-based materials.

\subsection{Magnetic Exchange}

Finally, we address the magnetic exchange in the Sm-Sm, Sm-Co, and Co-Co subsystems.
In the tetragonal lattice structure of \smco, the Sm and Co atoms both form two-dimensional square lattices parallel to the basal $ab$ plane, as shown in Fig. 1(a-b).
The Sm and Co layers alternate along the $c$ axis with a direct stacking structure, such that a Co atom sits directly between two Sm atoms.
The nearest neighboring Sm atoms in the $ab$ plane are separated by the $a$ lattice parameter of 4.5 \AA\xspace and contain an In(1) ion in the same plane, while the In(2) ions sit between the Sm and Co layers.
If the Sm and Co ions have a direct magnetic exchange pathway between them, it could also involve the In(2) sites, which could create a magnetic superexchange-like interaction.
An interaction involving In(2) is plausible because the wavefunction shapes of the Co $d_{xz}$/$d_{yz}$ orbitals lie closest to the In(2) sites, as shown in Fig. 1(a-b).
Depending on the details of hybridization with the In(2) $p_x, p_y,$ or $p_z$ orbitals, the magnetic superexchange mechanism could support multiple pathways with different ferromagnetic (FM) or antiferromagnetic (AF) exchange constants, in addition to the in-plane exchange pathway coupling Sm to Sm across the In(2) $p_x$ orbital.
Therefore, the In(2) site may provide another degree of freedom in tuning the magnetic exchange pathway between Sm and Co, and through it, the proportion of heavy quasiparticles entering the Fermi surface at the $\gamma_Z$ point versus those entering at $R$.
Since it was previously shown that hybridization between the rare earth and In(2) site is an important factor in controlling superconductivity in the \ceco-CeRhIn$_5$-CeIrIn$_5$ series \cite{WHHK10,WSHS15}, the complexity of the interactions that we observe in \smco may hold a key to understanding the origin of the ground state.
Specifically, the existence of competing magnetic exchange pathways involving the rare earth (R) sites, specifically, R-In(2)-R (in plane), R-Co-R (out of plane), and R-In(2)-Co-In(2)-R, may have consequences for how magnetic and/or superconducting ground states form in the 115s.

In addition to the RKKY mechanism, the role of direct magnetic exchange and magnetic superexchange can be generally important for $d$ and $f$ electrons in rare earth intermetallics.
For instance, in (La,Tb)Mn$_2$Si$_2$, it was proposed that an R-Mn-R interlayer superexchange mechanism is responsible for stabilizing AF order, while the Si atoms could act as mediators for superexchange between neighboring planes of Mn atoms along the $c$ axis \cite{KSGM18,KFSG20}.
In \smco, hybridization of the Sm $f$ and Co $d$ orbitals, possibly involving superexchange across the In(2) sites, suggests that these exchange mechanisms have more than one way to affect the magnetic Hamiltonian, particularly along the $c$-axis.
Such magnetic exchange pathways would therefore compete with the RKKY mechanism, and could play a destabilizing role in the magnetic ground state, particularly for interactions along the $c$ axis.
Competition between these magnetic interactions may therefore help explain the complicated magnetic phase diagram of \smco \cite{PJWR18}.
Therefore, our observation of flat $d$ bands interacting with states near the Fermi energy suggests a mechanism for FM and AF interactions to compete in \smco beyond the RKKY model, and also suggests that because of this competition, fine tuning by strong correlation effects among the $d$ states may have an outsize effect on the ultimate ground state of the material and the ordered structure of Sm local moments.

\subsection{Summary}

In summary, close agreement between ARPES measurements and DFT+U calculations shows that the band structure of \smco exhibits coupling of the localized Sm $f$ states at the Fermi level via hybridization with specific In $5p$ and Co $3d$ metallic states.
Our model shows that heavy quasiparticles containing Sm $f$ character can hybridize at multiple positions in the Brillouin zone, including a position supporting Sm-In(2)[$5p_x$] hybridization that is similar to previous results on \ceco, and a second position at the zone boundary supporting Sm-Co[$3d_{xz}$] hybridization (simultaneously with the equivalent orbitals $5p_y$ and $3d_{yz}$, respectively, along the other in-plane direction in the crystal).
Meanwhile, the Co $d$ orbitals exhibit a globally flat band ($\digamma$), which has a likely origin in geometric frustration.

The magnetic interactions arising from coupling of the itinerant quasiparticles to the Co layer, whether through direct hybridization (e.g. at $R$), through mean field interactions (because of a partly unfilled Co $d$ shell), or through superexchange interactions involving the In(2) sites, all point to the importance of the Co $d$ states in determining the ground state.
Since the $d$ shell is common to all ``115'' materials, electronic localization may play an overlooked role in the ground states found in this family, including unconventional superconductivity.
In \smco, the delocalized nature of Sm states at the hot spots shows how magnetic interactions can go beyond the typical RKKY mechanism, and may explain why the Sm $f$ states in \smco exhibit both a delocalized component and localized magnetic order.

\section{Methods}

\subsection{Sample growth}

\smco single crystals were grown using an indium flux method that we described previously \cite{sm115delocalization}.
Crystals were separated from the flux by centrifugation and later etched in hydrochloric acid.
The \smco samples exhibited a plate-like habit corresponding to the basal plane of the crystal.

\subsection{ARPES experiments}

Angle-resolved photoemission spectroscopy (ARPES) experiments were carried out at the soft-x-ray endstation \cite{adress1} of the ADRESS beamline \cite{adress2} at the Swiss Light Source, PSI, Switzerland.
Single crystals were mounted using the natural basal plane on copper plates using high-strength H20E silver epoxy.
Posts made of small stainless steel screws were attached onto the top surfaces of the samples with Torr-Seal epoxy.
The samples were cleaved in ultrahigh vacuum of about $10^{-10}$ mbar.
We used circularly polarized light with a beam spot of about 60 x 100 $\mu$m on the sample.
We determined the coupling of beam energy to the out-of-plane $k_z$ direction by scanning the beam energy over $h\nu=$480 to 700 eV, shown in Fig. 1(d).
We identified $h\nu=$ 602 eV as the energy where the in-plane iso-\eb maps correspond to the $\Gamma-X-M$ plane, and $h\nu=$ 564 eV as the energy corresponding to the $Z-R-A$ plane.
The use of soft-x-ray photon energies results in large photoelectron mean free paths, and the latter translates, by the Heisenberg relation, to sharp definition of the out-of-plane momentum $k_z$ \cite{Stro03,SSKM12}.
Maps as a function of two-dimensional in-plane momentum ($k_x$,$k_y$) were then collected at these energies to observe the Fermi surface and band structure, where we found broad similarity to published results of \ceco that use DFT \cite{CXNJ17,KFNM21} and dynamical mean-field theory \cite{JDAZ20}.

\subsection{DFT calculations}

All the density functional theory (DFT) calculations were carried out using the Quantum ESPRESSO
package \cite{N1,N2,N3}. Exchange and correlation effects were modelled using the PBEsol functional
\cite{N4}, augmented by a Hubbard U term to better describe the physics of the localized Sm $4f$ electrons.
The scalar-relativistic pseudopotentials are taken from the SSSP library \cite{N5}. To sample the Brillouin
zone, we used a 7x7x7 Monkhorst-Pack grid, a 10x10x10 grid was used for the calculation of the projected
density of states. The wave-functions (charge  densities and potentials) were expanded using a kinetic
energy cutoff of 70 Ry (560 Ry). We used the experimental lattice constants and internal atomic
positions \cite{N6,N7} throughout this work. The values of U used in this work (U=6.18 eV and U=6.26 eV
for Sm$^{3+}$ and Sm$^{2+}$ calculations, respectively) were computed fully ab-initio by using the linear
response approach \cite{N8} as implemented in Ref. \cite{N9,N10}. In order to ``nudge'' the self-consistent-field
calculations to converge to the Sm$^{3+}$ solution, we prepared the systems in initial configurations that
have 5 electrons in the Sm $f$ shell. Some of these initial configurations eventually converge to a
self-consistent solution that exhibit the desired $3^+$ oxidation state for the Sm atoms. At the level of DFT+U this
solution is always metastable, i.e. its energy is slightly higher then that of the Sm$^{2+}$ (6 electrons
in the Sm $f$ shell) solution. However it is the one that best reproduces the ARPES data.
Also, we did not use any +U corrections for the Co 3$d$ states as this deteriorates the agreement with the
ARPES data.
Finally, in the magnetic calculations including the $f$ electrons, the unfolding of the bands in the 7-atom
Brillouin zone was performed using the unfold-x code \cite{N11,PCAV21}.


\section{Data availability}

The data used to produce the simulation results presented in this work are available at the Materials Cloud Archive \cite{MC}.
Experimental datasets are available from the corresponding author on reasonable request.

\section{Acknowledgments}

This research was supported by the Swiss National Science Foundation (SNSF) under grant 200021\_184983 (M.K.), and by the NCCR MARVEL, a National Centre of Competence in Research, funded by the Swiss National Science Foundation (grant number 205602) (N.C.).
D.W.T. acknowledges funding from the European Union’s Horizon 2020 research and innovation programme under the Marie Skłodowska-Curie grant agreement No 884104 (PSI-FELLOW-III-3i).
F.A. acknowledges support by Swiss National Science Foundation Project No. 206312019002.

\section{Competing interests}

The authors declare no competing financial or non-financial interests.

\section{Author contributions}

D.W.T. and M.K. conceived the work and guided the project. D.J.G. and E.P. prepared samples, D.W.T. and F.A. prepared and conducted ARPES experiments with guidance from V.N.S., and N.C. performed density functional theory calculations. D.W.T, M.K., and N.C. prepared the manuscript with input from all authors.

\end{document}